\documentclass{iopart}
 \usepackage{graphicx}
 \usepackage{bm}

\begin{document}

 \title{Surface Enhanced van der Waals force}
 \author{Marin-Slobodan Toma\v s}
 \address{Rudjer Bo\v skovi\' c Institute, P. O. B. 180,
 10002 Zagreb, Croatia}
 \ead{tomas@thphys.irb.hr}

 \begin{abstract}
We consider the van der Waals interaction of an excited atom and a
ground state atom across a vacuum-medium interface under the
circumstances of the resonant coupling of the excited atom to a
surface polariton mode of the system. We demonstrate that the
nonretarded interaction potential between the atoms is in this
case the same as in an effective medium described by the average
dielectric function of the media in contact except for the
different local-field correction factor. The estimate performed
for the vacuum-sapphire interface shows that the nonretarded van
der Waals (atom*-atom) force can be at the surface mode resonance
enhanced by almost three orders of magnitude in comparison with
its free-space value. Owing to the local-field effect in the
present configuration this enhancement factor is larger then
previously estimated for atoms in front of the same medium by
(almost) an order of magnitude.
\end{abstract}

\section{Introduction}
It is well known that the decay rate and energy of an excited atom
(atom*) may be significantly affected near a dispersive surface
due to the resonant coupling of the atom to the surface polariton
mode(s) of the system \cite{CPS,FW,WS1,WS2,Bar,Fich,Buh,GoDu}.
Owing to the resonant energy shift of the atom, a strong
modification of the related atom*-surface force leading even to
the atom-surface repulsion has been predicted
\cite{CPS,WS2,Fich,Buh,GoDu} and observed \cite{Fail}. Based on
the Sherkunov formula for the interaction energy between an
excited and a ground-state atom \cite{YS}, we have recently
pointed out that the atom*-atom van der Waals potential/force is
also strongly enhanced near a dispersive surface whenever the
transition frequency of the excited atom coincides with the
frequency of a resonant mode of the system \cite{Tom08} .

Indeed, consider an excited ($e$) atom A and a ground-state ($g$)
atom B embedded in an inhomogeneous magnetoelectric system
described by the permittivity $\varepsilon({\bf r},\omega)$ and
permeability $\mu({\bf r},\omega)$. Assuming, for simplicity,
two-level isotropic atoms, the interaction potential between the
atoms is according to \cite{YS} [Eq. (68)] given by
\begin{numparts}
\label{UAB}
\begin{equation}
U_{AB}({\bf r}_A,{\bf r}_B)=U^{\rm or}_{AB}({\bf r}_A,{\bf
r}_B)+U^{\rm r}_{AB}({\bf r}_A,{\bf r}_B),\nonumber
\end{equation}
\begin{eqnarray}
\fl U_{AB}^{\rm or}({\bf r}_A,{\bf r}_B)&=&-\frac{\hbar}{2\pi
c^4}\int_0^\infty
\rmd\xi\xi^4\alpha^A_e(\rmi\xi)\alpha^B_g(\rmi\xi)
\Tr\left[\mathbf{G}({\bf r}_A,{\bf r}_B;\rmi\xi)\cdot
\mathbf{G}({\bf r}_B,{\bf r}_A;\rmi\xi)\right],\\
\fl U_{AB}^{\rm r}({\bf r}_A,{\bf r}_B)&=&-\frac{{\rm
Re}[\alpha^B_g(\omega_A)]\omega^4_A}{3c^4}|{\bf
d}^A_{eg}|^2\Tr\left[\mathbf{G}({\bf r}_A,{\bf
r}_B;\omega_A)\cdot\mathbf{G}^*({\bf r}_B,{\bf
r}_A;\omega_A)\right],
\end{eqnarray}
\end{numparts}
where $U_{AB}^{\rm or}$ and $U_{AB}^{\rm r}$ are off-resonant and
resonant part of the potential, respectively. In Eq. (\ref{UAB}),
$\alpha^X_{e(g)}(\omega)$ ($X=A,B$) are the atomic
polarizabilities whereas $\omega_A=(E_{e_A}-E_{g_A})/\hbar$ and
${\bf d}^A_{eg}=<e_A|{\bf d}^A|g_A>$ are, respectively, the
transition frequency and the dipole matrix element of atom $A$.
The dyadic $\mathbf{G}({\bf r},{\bf r'};\omega)$ is the classical
Green function for the system satisfying [$\mathbf{I}$ is unit
dyadic]
 \begin{equation}
 \left[\nabla\times\frac{1}{\mu({\bf r},\omega)}\nabla\times-
 \varepsilon({\bf r},\omega)\frac{\omega^2}{c^2}\mathbf{I}\cdot\right]
 \mathbf{G}(\omega;{\bf r},{\bf r'})
 =4\pi\mathbf{I}\delta({\bf r}-{\bf r'}),
 \label{GF}
 \end{equation}
with the outgoing wave condition at infinity. Poles of this Green
function in the complex $\omega$-plane correspond to frequencies
of the system resonant (polariton) modes. Accordingly, whenever
$\omega_A$ is close to the frequency $\omega_r$ of a resonant mode
the Green function behaves as $\mathbf{G}({\bf r}_A,{\bf
r}_B;\omega_A)\sim(\omega_A-\omega_r+i\gamma_r/2)^{-1}$ and, owing
to the factor $|G_{ij}({\bf r}_A,{\bf r}_B;\omega_A)|^2$, the van
der Waals potential is resonantly enhanced with respect to its
free-space value.

\begin{figure}[htb]
\begin{center}
\resizebox{8cm}{!}{\includegraphics{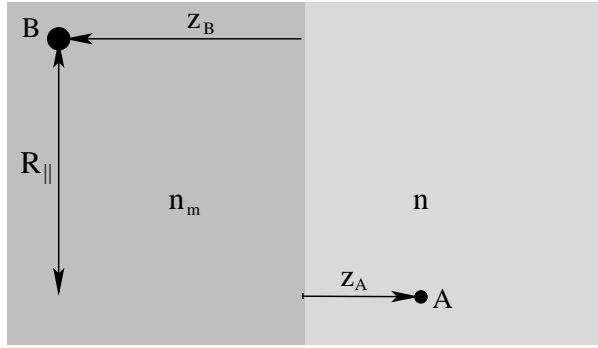}}
\end{center}
 \caption{Two atoms near an interface shown schematically.
 Media are described by (complex) refraction indexes
 $n(\omega)=\sqrt{\varepsilon(\omega)\mu(\omega)}$ and
 $n_m(\omega)=\sqrt{\varepsilon_m(\omega)\mu_m(\omega)}$.
\label{AB}}
\end{figure}

In order to verify the above considerations and estimate the
enhancement of the atom*-atom van der Waals interaction, in
\cite{Tom08} we have considered the interaction between the atoms
in front of the interface between two media under the
circumstances of the resonant coupling of the excited atom to the
surface polariton mode of the system. Here, we complement this
work, by considering the atom*-atom interaction across the
interface between two media, as depicted in Fig. \ref{AB}.

\section{Theory}

To account for the local-field effects, we adopt the Onsager model
and therefore assume small empty spherical cavities around the
atoms. Provided that $\omega_{\rm max} R_{A(B)}/c\ll 1$, with
$R_{A(B)}$ being cavity radii and $\omega_{\rm max}>\omega_{A(B)}$
an effective cutoff frequency in (\ref{UAB}), the Green function
for this system can [to the order of $\omega R_{A(B)}/c$] be
written as \cite{Sam}
\begin{equation}
\label{Gl} \mathbf{G}({\bf r}_A,{\bf r}_B;\omega)=D(\omega)
\tilde{\mathbf{G}}({\bf r}_A,{\bf r}_B;\omega) D_m(\omega),
\end{equation}
where
\begin{equation}
D(\omega)=\frac{3\varepsilon(\omega)}{2\varepsilon(\omega)+1},\;\;\;
D_m(\omega)=
\frac{3\varepsilon_m(\omega)}{2\varepsilon_m(\omega)+1},
\end{equation}
and (${\bf R}_\parallel\equiv {\bf r}_{A\parallel}- {\bf
r}_{B\parallel}$)
\[\tilde{\mathbf{G}}({\bf r}_A,{\bf r}_B;\omega)=\int\frac{{\rm
d}^2{\bf k}}{(2\pi)^2}e^{i{\bf k}\cdot{\bf
R}_\parallel}\tilde{\mathbf{G}}(z_A,z_B;\omega,{\bf k}),\]
\begin{equation}
\fl \tilde{\mathbf{G}}(z_A,z_B;\omega,{\bf k}) =2\pi
i\frac{\mu}{\beta} \left[t^p\frac{\beta\hat{\bf k}-k\hat{\bf
z}}{n\omega/c}\frac{\beta_m\hat{\bf k}-k\hat{\bf z}}{n_m\omega/c}
+t^s\hat{\bf k}\times\hat{\bf z}\hat{\bf k}\times\hat{\bf
z}\right]e^{i\beta z_A-i\beta_mz_B} \label{G}
\end{equation}
is the Green function for the system unperturbed by the Onsager
cavities \cite{Tom95}. Here $\beta=\sqrt{(n\omega/c)^2-k^2}$ and
$\beta_m=\sqrt{(n_m\omega/c)^2-k^2}$ are the perpendicular wave
vectors in the media whereas
\begin{equation}\label{tps}
t^p(k,\omega)=\sqrt{\frac{\varepsilon\mu_m}{\varepsilon_m\mu}}
\frac{2\varepsilon_m\beta}{\varepsilon_m\beta+\varepsilon\beta_m}\;\;\;{\rm
and}\;\;\; t^s(k,\omega)=\frac{2\mu_m\beta}{\mu_m\beta+\mu\beta_m}
\end{equation}
are the Fresnel transmission coefficients for the interface.
Considering the situation where the atomic distances from the
interface between the media are small compared to $c/\omega_{\rm
max}$ (but larger than $R_{A(B)}$) we can neglect the retardation
of the electromagnetic field. The Green function in the
nonretarded approximation $\tilde{\mathbf{G}}_{\rm nr}({\bf
r}_A,{\bf r}_B;\omega)$ is easily obtained using the prescription
\cite{Tom95}
\[\tilde{\mathbf{G}}_{\rm nr}({\bf r}_A,{\bf r}_B;\omega)
=\frac{c^2}{\omega^2}\lim_{c\rightarrow\infty}
\frac{\omega^2}{c^2}\tilde{\mathbf{G}}({\bf r}_A,{\bf
r}_B;\omega)\] and we find
\begin{equation}
\tilde{\mathbf{G}}_{\rm nr}({\bf r}_A,{\bf r}_B;\omega)
=\frac{c^2}{\overline{\varepsilon}(\omega)\omega^2}\int\frac{{\rm
d}^2{\bf k}}{2\pi}k(i\hat{\bf k}-\hat{\bf z})(i\hat{\bf
k}-\hat{\bf z})e^{i{\bf k}\cdot{\bf R}_{\parallel}-k(z_A-z_b)},
\end{equation}
where
\begin{equation}
\overline{\varepsilon}(\omega)=
\frac{\varepsilon(\omega)+\varepsilon_m(\omega)}{2}
\end{equation}
is the average dielectric function of the two media in contact. As
seen, this Green function is the same as the nonretarded Green
function in an infinite medium described by the dielectric
function $\overline{\varepsilon}(\omega)$. Accordingly, going back
to the $({\bf R},\omega)$-space recalling that
\begin{equation}
\frac{1}{R}=\int\frac{{\rm d}^2{\bf k}}{2\pi k}e^{i{\bf
k}\cdot{\bf R}_{\parallel}-k|z_A-z_b|},\;\;\;{\bf R}={\bf
r}_A-{\bf r}_B,
\end{equation}
we finally arrive at
\begin{equation}
\label{Gnr} \mathbf{G}_{\rm nr}({\bf r}_A,{\bf
r}_B;\omega)=\frac{D(\omega)D_m(\omega)}{(c/\omega)^2\overline{\varepsilon}(\omega)}
\frac{3{\bf R}{\bf R}-R^2\mathbf{I}} {R^5}.
\end{equation}

Inserting the above Green function in Eq. (\ref{UAB}) and
neglecting (small) off-resonant part of the potential, we find
that $U_{AB}$ is given by
\begin{equation}
\label{UABf} U_{AB}({\bf r}_A,{\bf r}_B)=-\frac{2{\rm
Re}[\alpha^B_g(\omega_A)]|{\bf
d}^A_{eg}|^2|D(\omega_A)|^2|D_m(\omega_A)|^2}
{|\overline{\varepsilon}(\omega_A)|^2R^6}.
\end{equation}
and is enlarged by a factor of
\begin{equation}
g(\omega_A)=\left|\frac{18\varepsilon(\omega_A)\varepsilon_m(\omega_A)}
{[\varepsilon(\omega_A)+\varepsilon_m(\omega_A)]
[2\varepsilon(\omega_A)+1][2\varepsilon_m(\omega_A)+1]}\right|^2
\end{equation}
with respect to its free-space value. Evidently, the potential is
resonantly enhanced whenever $\omega_A$ coincides with the
frequency of an interface (surface) mode or with the frequency of
a resonant mode of the Onsager cavities. We also observe that in
this configuration, as long as the atoms are close to the
interface, $U_{AB}({\bf r}_A,{\bf r}_B)$ is isotropic and depends
only on the distance between the atoms in contrast to the
situation when both atoms are on the same side of the interface
\cite{Tom08}. Accordingly, the van der Waals forces on atoms ${\bf
F}_X=-\nabla_XU_{AB}({\bf r}_A,{\bf r}_B)$ obey the third Newton
law
\begin{equation}
{\bf F}_A({\bf r}_A,{\bf r}_B)=-\frac{12{\rm
Re}[\alpha^B_g(\omega_A)]|{\bf d}^A_{eg}|^2g(\omega_A)}
{R^7}\hat{\bf R}=-{\bf F}_B({\bf r}_B,{\bf r}_A)
\end{equation}
despite the inhomogeneity of the system. Of course, this is not
the case with total vacuum-field forces on atoms since they also
include the respective Casimir-Polder forces.

\section{Discussion}
In order to compare the surface enhanced van der Waals interaction
between the atoms in the present configuration and previously
considered one \cite{Tom08}, we consider the same interface and
therefore assume the atom $A$ located in the vacuum
[$n(\omega)=1$] and the atom $B$ embedded in a nonmagnetic medium
that is around a resonance at $\omega_T$ described by the
dielectric function \cite{Bar}
\begin{equation}
\label{em}
\varepsilon_m(\omega)=\eta+\frac{(\varepsilon_0-\eta)\omega_T^2}
{\omega_T^2-\omega^2-\rmi\omega\Gamma},
\end{equation}
where $\eta$ and $\varepsilon_0\equiv\varepsilon(0)$ are
background and static dielectric constant, respectively, whereas
$\Gamma$ is the corresponding damping. Quantity
$(\varepsilon_0-\eta)\omega_T^2$, denoted as $\eta\omega_P^2$ in
\cite{Tom08}, is a measure of the coupling between the medium
polarization and the field. The inverse of the average dielectric
function of the system can therefore be written as
\begin{equation}
\label{eav}
\frac{1}{\overline{\varepsilon}(\omega)}=\frac{2}{\eta+1}\left(1-
 \frac{\varepsilon_0-\eta}{\varepsilon_0+1}
 \frac{\omega_S^2}{\omega_S^2-\omega^2-\rmi\omega\Gamma}\right),
\end{equation}
where
\begin{equation}
\label{wS}
 \omega_S=\sqrt{\frac{\varepsilon_0+1}{\eta+1}}\omega_T
\end{equation}
is the surface-mode frequency. In conjunction with Eq.
(\ref{UABf}), this equation explicitly demonstrates resonant
enhancement of $U_{AB}$ when $\omega_A\sim\omega_S$. We note,
however, that $D_m(\omega)$ also exhibits a resonant structure.
Indeed, with Eq. (\ref{em}) this function reads
\begin{equation}
\label{Dm} D_m(\omega)=\frac{3}{2\eta+1}\left(\eta+
 \frac{\varepsilon_0-\eta}{2\varepsilon_0+1}
 \frac{\omega_C^2}{\omega_C^2-\omega^2-\rmi\omega\Gamma}\right),
\end{equation}
where
\begin{equation}
\omega_C= \sqrt{\frac{2\varepsilon_0+1}{2\eta+1}}\omega_T
\end{equation}
is the surface-mode frequency of the Onsager cavity. Since
$\omega_C$ is pretty close to $\omega_S$ the enhancements of
$U_{AB}$ due to two resonances (partially) combine. Indeed, for
small damping we find that $D_m(\omega_S)\simeq 3$ so that the
peak value of the surface enhanced van der Waals potential/force
given by
\begin{equation}
g(\omega_S)\simeq
\frac{4(\varepsilon_0-\eta)^2}{(\eta+1)^2(\varepsilon_0+1)^2}
\frac{\omega_S^2}{\Gamma^2}|D_m(\omega_S)|^2\label{gs}
\end{equation}
is enlarged by (almost) nine times due to the local-field effect.

\begin{figure}[htb]
\begin{center}
\resizebox{8cm}{!}{\includegraphics{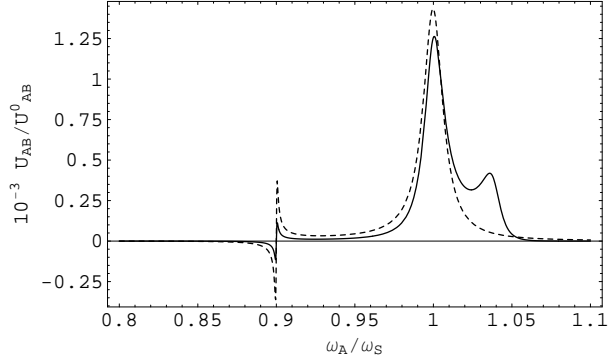}}
\end{center}
 \caption{Surface enhanced van der Waals potential with (full line)
 and without (dashed line) local-field correction factor (in the former
 case the potential is additionally scaled by a factor of 10).
 Relevant medium parameters are $\eta=2.71$,
 $\varepsilon_0=6.57$, $\Gamma=0.015\omega_S$ and $\omega_S=
 1.54\times 10^{14}{\rm s}^{-1}$ \cite{GoDu}. The transition
 frequency and linewidth of
 atom B are $\omega_B=0.9\omega_S$ and $\gamma_B=10^{-3}\omega_S$,
 respectively. \label{Upo}}
\end{figure}

We illustrate these considerations in Fig. \ref{Upo} where we have
plotted $U_{AB}^{\rm r}$ with and without local- field correction
factor as a function of $\omega_A$. Note that the local-field
corrected potential (full line) is additionally scaled by a factor
of 10. To obtain these curves we have let in Eq. (\ref{UABf})
\begin{equation}
\label{alpha} \alpha^B_g(\omega)=\alpha^B_g(0)\frac{\omega_B^2}
{\omega_B^2-\omega^2-\rmi\omega\gamma_B}
\end{equation}
and expressed the potential in units $U_{AB}^0=2|{\bf
d}^A_{eg}|^2\alpha^B_g(0)/R^6$. Medium parameters are taken the
same as in \cite{Tom08} and correspond to sapphire around the
surface polariton resonance at $\lambda_S= 12.21 {\rm \mu m}$
\cite{GoDu} whereas parameters of the atom B are chosen quite
arbitrarily. We see that, besides a resonant structure at
$\omega_B$ as would exist in the free-space, the potential
exhibits also resonances at $\omega_S$ and
$\omega_C=1.04\;\omega_S$ corresponding to the surface and
local-field enhanced interaction between the atoms, respectively.
As mentioned, owing to the proximity of these two resonances each
peak is actually due to the combined surface and local-field
effects. Thus, for the local-field corrected surface enhanced
potential $g(\omega_S)=2947.6$ whereas for the potential without
the local-field correction the enhancement factor is 8.8 times
smaller, $g(\omega_S)=336.2$,which is of the same order as
$g(\omega_S)$ found in \cite{Tom08} for the case when the atoms
are in vacuum close to the sapphire surface. Similarly, we find
for the total enhancement factor at the Onsager cavity resonance
$g(\omega_C)=1217.9$ whereas the enhancement factor without the
contribution of the surface-mode resonance at $\omega_S$ is
$g(\omega_C)=79.3$.

\section{Summary}
In this work, we have considered the local-field corrected van der
Waals interaction between an excited and a ground state atom
across the interface between two media. We have demonstrated that
the (nonretarded) atom*-atom potential is of the same form as in a
homogeneous medium described by the average dielectric function of
the two media in contact. Accordingly, as a function of the
transition frequency of the excited atom, the potential generally
exhibits resonances associated with the interface (surface) modes
and with modes of small spherical cavities around the atoms
figuring in the Onsager model for the local field. Numerical
calculation performed for the vacuum-sapphire system reveals that
in such systems, owing to the close surface and the Onsager cavity
resonances, the van der Waals potential can be at the
(vacuum-medium) surface mode frequency enhanced by over three
orders of magnitude of which (nearly) one order of magnitude is
due to the enhancement of the local-field correction factor for
the medium.

\ack This work was supported by the Ministry of Science, Education
and Sport of the Republic of Croatia under contract No.
098-1191458-2870.


\section*{References}

\begin{thebibliography}{00}

\bibitem{CPS}Chance R R, Prock A and Silbey R 1978 {\it
Advances in Chemical Physics} {\bf 37}, 1
\bibitem{FW}Ford G W and Weber W H 1984 {\it Phys. Rep.} {\bf 113} 195
\bibitem{WS1}Wylie J M and Sipe J E 1984 {\it Phys. Rev. A}
{\bf 30} 1185
\bibitem{WS2}Wylie J M and Sipe J E 1985 {\it Phys. Rev. A}
{\bf 32} 2030
\bibitem{Bar}Barton G 1997 {\it Proc. Roy. Soc. Lond. A} {\bf 453} 2461
\bibitem{Fich}Fichet M, Schuller F, Bloch D and Ducloy M 1995
{\it Phys. Rev. A} {\bf 51} 1553
\bibitem{Buh}Buhmann S Y, Knoll L, Welsch D-G and
Dung H T 2004 {\it Phys. Rev. A} {\bf 70} 052117
\bibitem{GoDu}Gorza M-P and Ducloy M 2006 {\it Eur. Phys. J. D} {\bf 40}
343
\bibitem{Fail}Failache H, Saltiel S, Fichet M, Bloch D and
Ducloy M 1999 {\it Phys. Rev. Lett.} {\bf 83} 5467
\bibitem{YS}Sherkunov Y 2007 {\it Phys. Rev. A} {\bf 75} 012705
\bibitem{Tom08}M. S. Toma\v{s}, 2008 {\it J. Phys. A: Mathematical and
Theoretical} {\bf 41} 164020
\bibitem{Sam}Sambale A, Buhmann S Y, Welsch D-G, and
Toma\v{s} M S, 2007 {\it Phys. Rev. A} {\bf 75} 042109
\bibitem{Tom95}Toma\v s M S 1995 {\it Phys. Rev. A} {\bf 51} 2545


\end{thebibliography}
\end{document}